\pdfoutput=1
\documentclass{PoS}

\usepackage[utf8]{inputenc}
\usepackage{amsmath}
\usepackage{url}

\newcommand{\Tr}{\mathrm{Tr}\,}

\title{Scalar QCD at nonzero density}

\ShortTitle{Scalar QCD at nonzero density}

\author{Falk Bruckmann and \speaker{Jacob Wellnhofer} \\
       Universität Regensburg, Universitätsstraße 31, 93053 Regensburg, Germary\\
			 E-mail: \email{falk.bruckmann@ur.de},\email{jacob.wellnhofer@ur.de}}

\abstract{
	We study scalar QCD at nonzero density in the strong coupling limit. It has a sign problem which looks structurally similar to the one in QCD. We show first data for the reweighting factor. After introducing dual variables by integrating out the SU(3) gauge links, we find that at least 3 flavors are needed for a nontrivial dependence on the chemical potential. In this dual representation there is no sign problem remaining. The dual variables are partially constrained, thus we propose to use a hybrid approach for the updates: For unconstrained variables local updates can be used, while for constrained variables using updates based on the worm algorithm is more promising.
}

\FullConference{34th annual International Symposium on Lattice Field Theory\\
         24-30 July 2016\\
         University of Southampton, UK}

\begin{document}

\section{Introduction}
Lattice QCD is one of the most important tools for studying the 
nonperturbative as well as thermodynamic aspects of QCD from first principles. 
However, if we introduce a chemical potential $\mu$ in order to explore 
the phase-diagram of QCD at nonzero density, 
the standard approach fails due to the sign problem, that is, 
the weights of the gauge configurations (having integrated out the quarks) become complex and 
therefore ill-suited for importance sampling in Monte-Carlo simulations. 

There have been many proposals to remedy this shortcomming. 
Standard reweighting techniques fail because the 
reweighting factor rapidly approaches zero in the 
interesting regime around the critical chemical 
potential $\mu_c$. 
Another proposal has been the MDP-formulation (Monomer-Dimer-Polymer)
of strong coupling QCD \cite{ROSSIWOLFF,KARSCHMUTTER}. 
There first the gauge links  
and after that the fermion fields
are integrated out. 
This leaves a constrained spin system of 
occupation numbers or dual variables. 
However, even then there is still a sign problem 
remaining, even though it is rather mild \cite{FROMM}. 

In these proceedings we consider a scalar version of QCD in the strong coupling limit (scSQCD), 
where instead of fermionic fields we use complex scalar fields. 
We can couple a chemical potential to the conserved charge of the scalar fields, 
which also results in a complex action. 
We dualize the theory in a similar way to the MDP-formulation, 
which solves the sign problem. 
Afterwards we discuss the diagrammatic representation of the dual theory 
and propose a simulation strategy for it. 

\section{Strong coupling scalar QCD}
The action for this model reads
    \begin{align}
        S
        &=
        \sum_{x}
				\sum_{f=1}^{N_f}
        \left(
				\sum_{\nu=1}^{d}
        \left(
        e^{\mu\delta_{\nu,\hat 0}}\phi_x^{(f)\dagger} U_{x,\nu} \phi_{x+\hat\nu}^{(f)}
        +
        e^{-\mu\delta_{\nu,\hat 0}}\phi_{x+\hat\nu}^{(f)\dagger} U_{x,\nu}^\dagger \phi_{x}^{(f)}
        -2 \phi_x^{(f)\dagger}\phi_x^{(f)}
        \right)
        -m^2 \phi_x^{(f)\dagger}\phi_x^{(f)}
        \right)
    \end{align}
where $\phi$ is the complex scalar field, 
$U$ the SU(3) gauge field, $N_f$ the number of flavors and $d$ the 
number of spacetime dimensions. 

\begin{figure}[p]
\begin{center}
    \includegraphics[scale=1]{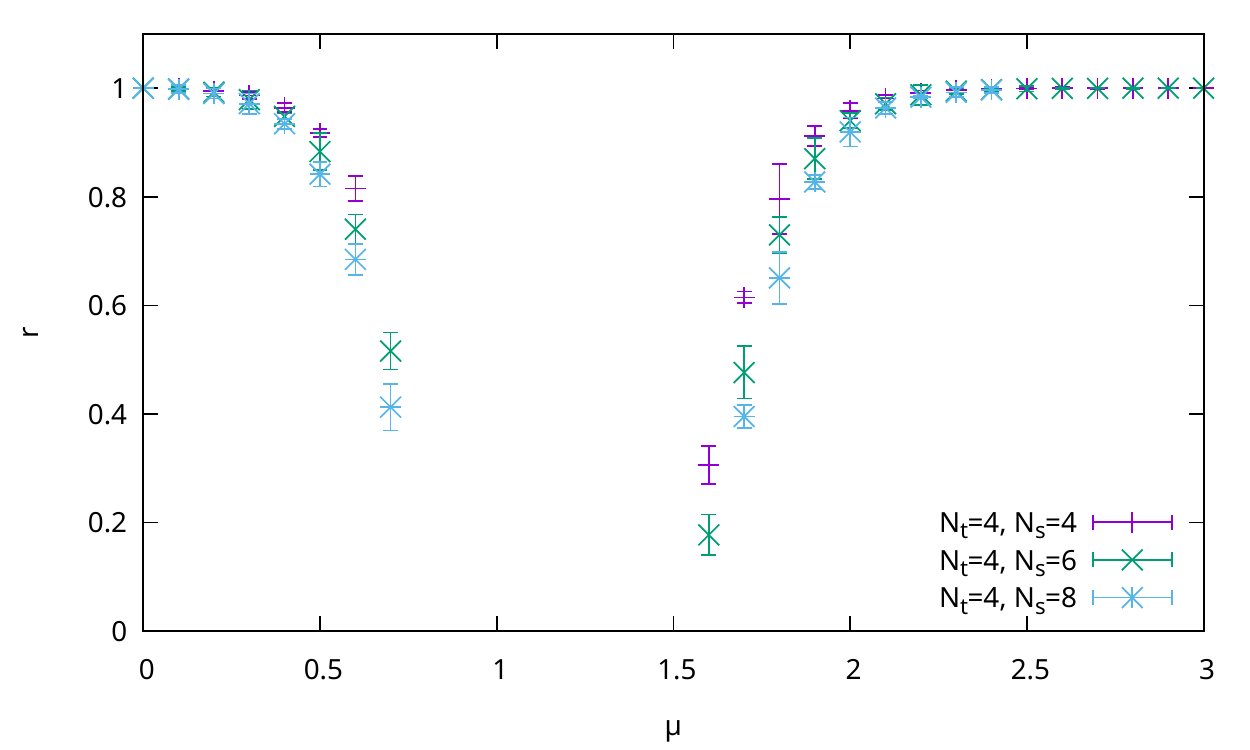}
\end{center}
\caption{
	The reweighting factor is plotted as a function of $\mu$. 
	The data is from 10000 configurations and we set $N_f=3, m = 0.1$.
	For $\mu \in [0.7,1.6]$ the autocorrelationtime becomes so large 
	that neither the errors nor the mean values are reliable any more. 
	Therefore we have omitted the data points in this region. 
	}
\label{fig:rewfac}
\end{figure}

\begin{figure}[p]
\begin{center}
    \includegraphics[scale=1]{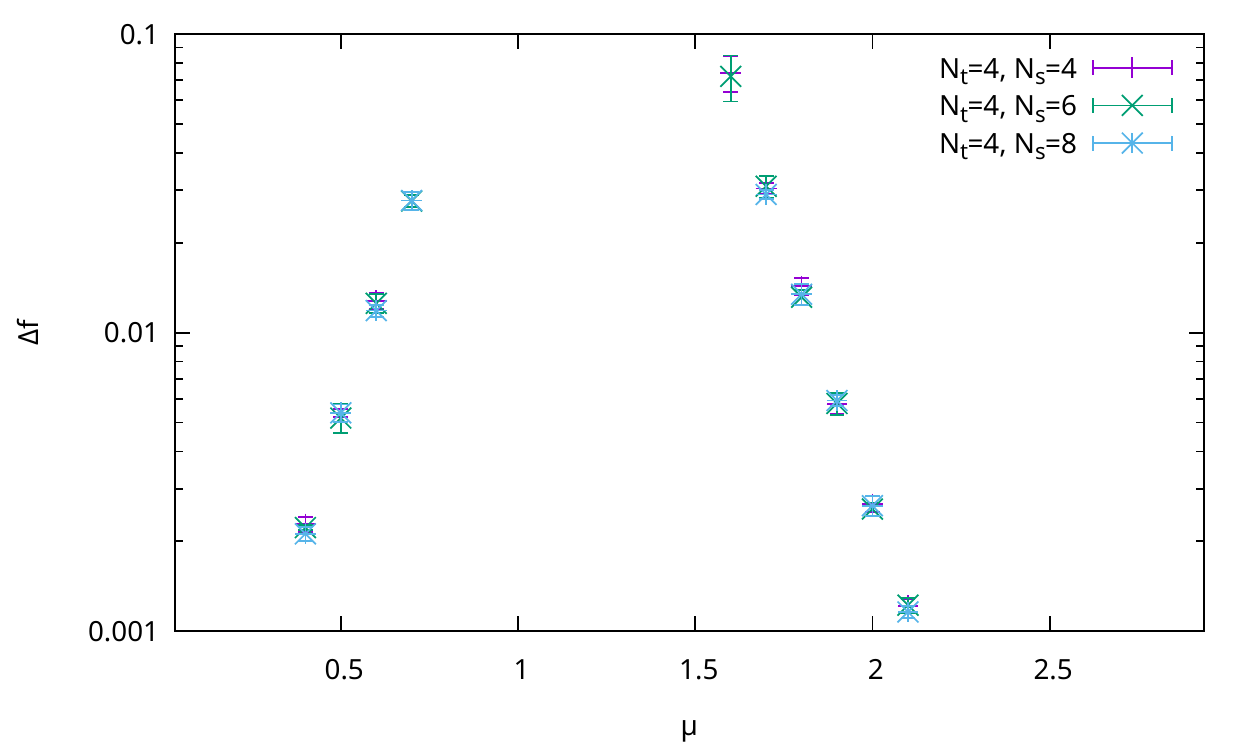}
\end{center}
\caption{
	The free energy density $\Delta f$, $r = e^{-\Delta f V}$, 
	is plotted as a function of $\mu$. 
	A naive extrapolation yields a crossing point at $\mu_\times = 1.172(2)$, 
	with a free energy density of $\Delta f_\times = 2.04(7)$. 
	}
\label{fig:deltaf}
\end{figure}

For $\mu\neq 0$ the action becomes complex and we end up with a sign problem. 
In figure \ref{fig:rewfac} the (phase-quenched) reweighting factor, 
$r = \langle(|\det M| / \det M)^{N_F}\rangle$,
is plotted as a function 
of the chemical potential $\mu$. Above $\mu = 0.7$ it decreases rapidly. 
Figure \ref{fig:deltaf} shows that $r = e^{-\Delta f V}$ 
obeys the correct volume dependence, where $\Delta f$ is the free energy density. 
Further we note that $\Delta f$ seems to have an exponential increase/decrease with $\mu$. 
A naive extrapolation gives a maximal free energy density of $\Delta f(\mu_\times) = 2.04(7)$ at 
$\mu_\times = 1.172(2)$. 
This means that the reweighting factor becomes smaller than $10^{-14}$ already for a $4\times 4$ lattice. 
Therefore it is unfeasible to simulate the region around $\mu_c$ by reweighting\footnote{
For large $\mu \gg \mu_c$ the reweighting factor becomes well behaved again.
}. 

\section{Dualization}
In order to proceed to dualize the theory, we rewrite the action:
\begin{align}
        S
        &=
        \sum_{x}
        \left(
        \sum_\nu
        \Tr
        \left(
        J_{x,\nu}U_{x,\nu}
        +
        K_{x,\nu}U_{x,\nu}^\dagger
        \right)
				-
        \sum_{f}
        \left(
				(2d +m^2) \phi_x^{(f)\dagger}\phi_x^{(f)}
        \right)
        \right)
        \\
        J_{x,\nu}
        &=
        e^{\mu\delta_{\nu,\hat 0}}
        \sum_f
        \phi_{x+\hat\nu}^{(f)}\phi_x^{(f)\dagger} 
				\hspace{3em}
				\mathrm{forward\ hopping}
				\label{eq:forwardhop}
        \\
        K_{x,\nu}
        &=
        e^{-\mu\delta_{\nu,\hat 0}} 
        \sum_f
        \phi_{x}^{(f)}\phi_{x+\hat\nu}^{(f)\dagger} 
				\hspace{2.7em}
				\mathrm{backward\ hopping}
				\label{eq:backwardhop}
\end{align}
For the partition function $Z$ we have to integrate over 
the gauge field $U$ as well as the scalar fields $\phi$:
\begin{align}
	Z
	&=
	\int \mathcal{D}\phi\mathcal{D}\phi^\dagger
	\int\mathcal{D}U\,
	e^{S}.
\end{align}
The SU(3) integral at a single bond $(x,\nu)$ can be turned into a five-fold sum, see, e.g., \cite{SKAND}:
            \begin{align}
							\label{eq:SKAND}
                &\int_{\mathrm{SU(3)}}dU\, 
                \exp(\Tr(JU + KU^\dagger))
                =
								\sum_{j,k,l,n,\bar n = 0}^{\infty}
                \frac{X^j}{j!} 
                \frac{Y^k}{k!} 
                \frac{Z^l}{l!} 
                \frac{\Delta^{n}}{n!} 
                \frac{\bar\Delta^{\bar n}}{\bar n!} 
                \frac{2}{f^{(1)}! f^{(2)}!} 
            \end{align}
where $f^{(1)}$ and $f^{(2)}$ are shorthands for
						\begin{align}
								f^{(1)}
								= k+2l+n+\bar n + 1
								\hspace{3em}
                f^{(2)} = j+2k+3l+n+\bar n +2, 
						\end{align}
and $X,Y,Z,\Delta,\bar\Delta$ are functions of $J,K$:
						\begin{align}
                X 
								&= \Tr(KJ)
								&
                Y
								&=
                \frac{1}{2} \left[ X^2 - \Tr((KJ)^2) \right]
								&
                Z
								&=
                \det(KJ)
								\\
                \Delta
								&= \det J
								&
                \bar\Delta
								&= \det K
						\end{align}
We can apply \eqref{eq:SKAND} to each bond separately 
because the SU(3) integrals factorize. 
Then the partition function becomes
    \begin{align}
			\label{eq:partition_function_dual}
        Z
        &=
        \sum_{\{j,k,l,n,\bar n\}}
				\int
        \mathcal{D}\phi
        \mathcal{D}\phi^\dagger
        \rho(|\phi|)
        \prod_{x,\nu}
        2 \frac{X^{j_{x,\nu}}_{x,\nu}Y^{k_{x,\nu}}_{x,\nu}Z^{l_{x,\nu}}_{x,\nu}\Delta^{n_{x,\nu}}_{x,\nu}\bar\Delta^{\bar n_{x,\nu}}_{x,\nu}}
        {j_{x,\nu}!k_{x,\nu}!l_{x,\nu}! n_{x,\nu}!\bar n_{x,\nu}! f^{(1)}_{x,\nu}! f^{(2)}_{x,\nu}!} 
				\\
				\rho(|\phi|)
				&=
				\prod_{x,f} e^{-(2d+m^2)|\phi^{(f)}_x|^2}
    \end{align}
Thus we are left with the integration over the scalar fields, 
which are gaussian distributed, and the summation over configurations
of dual variables $j,k,l,n,\bar n$. 

The functions $X,Y,Z$ only depend on the matrix $KJ$. 
In our case the $\mu$ dependence cancels, 
cf. \eqref{eq:forwardhop} and \eqref{eq:backwardhop},
and we have $KJ = J^\dagger J$, 
which is a positive operator, and hence 
$X,Y,Z$ are positive as well\footnote{
Note that in particular $Y$ is also positive. 
}. 
However, $\Delta, \bar\Delta$ are complex in general. 
One can work out that
\begin{align}
	\Delta_{x,\nu}
	&=
	e^{3\mu\delta_{\nu,0}}
	\sum_\sigma
	\phi^{(\sigma(1))}_{x+\hat\nu}
	\cdot
	\left(
	\phi^{(\sigma(2))}_{x+\hat\nu}
	\times
	\phi^{(\sigma(3))}_{x+\hat\nu}
	\right)
	\left(
	\phi^{(\sigma(1))}_{x}
	\cdot
	\left(
	\phi^{(\sigma(2))}_{x}
	\times
	\phi^{(\sigma(3))}_{x}
	\right)
	\right)^*
	\\
	\bar\Delta_{x,\nu}
	&=
	e^{-3\mu\delta_{\nu,0}}
	\sum_\sigma
	\phi^{(\sigma(1))}_{x}
	\cdot
	\left(
	\phi^{(\sigma(2))}_{x}
	\times
	\phi^{(\sigma(3))}_{x}
	\right)
	\left(
	\phi^{(\sigma(1))}_{x+\hat\nu}
	\cdot
	\left(
	\phi^{(\sigma(2))}_{x+\hat\nu}
	\times
	\phi^{(\sigma(3))}_{x+\hat\nu}
	\right)
	\right)^*
\end{align}
where $\sigma$ runs over all maps 
$\sigma: \{1,2,3\}\to \{1,\dots, N_f\}$
with $\sigma(1) < \sigma(2) < \sigma(3)$.
This means that for $N_f < 3$ there is 
no $\mu$ dependence since then $\Delta \equiv \bar\Delta \equiv 0$, 
see also \cite{WOLFF}.
In the following we restrict ourselves to the first nontrivial 
case, $N_f = 3$. However, the result is valid also for $N_f>3$. 

To tackle the remaining sign problem, 
note that the partition function involves 
gaussian integrations of the form 
\begin{align}
	\int d\phi d\phi^*\, e^{-|\phi|^2}
	(\phi)^a (\phi^*)^{ b}
	&\sim
	\delta_{ab}
	\label{eq:constraint}
\end{align}
at each site for each flavor and color. 
Thus the only contribution to the partition 
function comes from terms where the 
power of $\phi$ at a site matches that of $\phi^*$. 

$X,Y,Z$ satisfy the constraint at a single bond, 
however, $\Delta, \bar\Delta$ do not. 
From \eqref{eq:constraint} one can see that 
only closed loops of $\Delta, \bar\Delta$ satisfy the constraint. 
For such a closed loop $C$ the weight is proportional to
\begin{align}
	w(C)
	&\sim
	e^{3N_t\mu w_t}
	\prod_{(x,\hat\nu) \in C}
	\phi^{(1)}_{x}
	\cdot
	\left(
	\phi^{(2)}_{x}
	\times
	\phi^{(3)}_{x}
	\right)
	\left(
	\phi^{(1)}_{x+\hat\nu}
	\cdot
	\left(
	\phi^{(2)}_{x+\hat\nu}
	\times
	\phi^{(3)}_{x+\hat\nu}
	\right)
	\right)^*
	\\
	&=
	e^{3N_t\mu w_t}
	\prod_{x \in C}
	\left |
	\phi^{(1)}_{x}
	\cdot
	\left(
	\phi^{(2)}_{x}
	\times
	\phi^{(3)}_{x}
	\right)
	\right |^2
	\ge 0,
	\label{eq:loop_positive}
\end{align}
where $w_t$ is the winding number in the temporal direction.
So we have to consider only closed loop configurations in the partition function. 
Then there is no sign problem remaining even at $\mu \neq 0$\footnote{
Note that this only applies to real $\mu$; for complex values of $\mu$ 
a sign problem reappears.
}.

\section{Discussion and Conclusion}
\begin{figure}[h]
\begin{center}
	\includegraphics{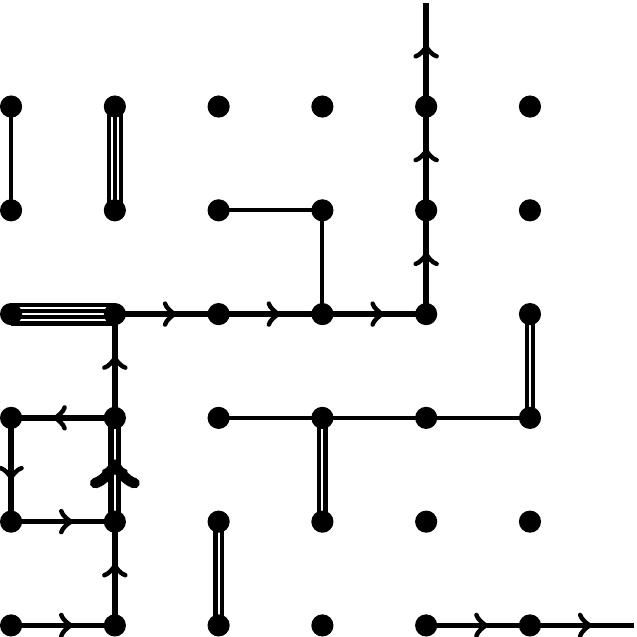}
\caption{
	A sample configuration of the dual variables.
	Bonds can be occupied arbitrarily with the 
	dual variables $j,k,l,n,\bar n$, 
	the only constraint being that the $n,\bar n$
	have to form closed loops. Since these loops 
	are directional, they are marked with arrows. 
	Note that in constrast to the MDP-formulation 
	there is also no restriction of \emph{baryonic}
	bonds ($n,\bar n$) being disjunct from \emph{mesonic}
	bonds ($j,k,l$) 
	and sites without occupied bonds are possible as well. 
	Also there is no restriction on the occupation numbers, 
	they can, in principle, be arbitrarily large. 
	However, large values are surpressed by the 
	factorials in the denominator of eq. (3.9).}
\label{fig:sample_config}
\end{center}
\end{figure}
The configurations of the dual variables can be represented diagrammatically 
in a similar way to the MDP-formulation, see figure \ref{fig:sample_config}.
We may call closed loops of $n,\bar n$ \emph{baryonic}, 
since they are directional, and couple to the chemical potential $\mu$. 
The other dual variables $j,k,l$ we call \emph{mesonic}. 
A notable difference to the MDP-formulation is that in our case 
there is no restriction of mesonic bonds being disjunct from baryonic ones. 
Also, since we are in a bosonic theory, there is no restriction on 
the dual variables. In principle they can run from $0$ to $\infty$, 
however, large values are strongly suppressed by the factorials in the 
denomitator of \eqref{eq:partition_function_dual}.

From the derivation of the original MDP-formulation it is evident 
that there the remaining sign problem stems from the fermionic nature of the fields. 
In particular, there the sign problem comes from 
the anticommutation rules, 
the staggered phases, 
the backward hoppings, 
and the antiperiodic boundary conditions in the time direction. 
In the scalar case all these causes are absent, 
and as we have seen this results in a dual theory that has no sign problem. 

For the simulation of this model we propose a hybrid strategy. 
The dual variables $j,k,l$ can be updated via a local metropolis step, 
which involves an evaluation of the functions $X,Y,Z$ on the 
corresponding bond, as $X,Y,Z$ are all positive.  
The $n,\bar n$ can be updated using a worm-type of algorithm. 
To that end one can use the fact that for a closed loop the 
contribution of each site is positive, see eq. \eqref{eq:loop_positive},  
and one can use a heat-bath method to decide which way to go with the worm.
Finally the (gaussian distributed) scalar fields $\phi, \phi^\dagger$ can also be 
updated via a local metropolis step by evaluating $X,Y,Z$ on the adjacent bonds 
as well as the contribution of the closed $n,\bar n$-loops to that particular site, 
cf. eq. \eqref{eq:loop_positive}. 
We leave numerical simulations which explore the phase diagram 
of this model for future publications. 

We thank Jacques Bloch for helpful discussions. 
This work is supported by the DFG-grants {BR 2872/6-1 \& 2872/7-1}.

\end{document}